\documentclass[fleqn,usenatbib]{mnras}

\usepackage[T1]{fontenc}
\usepackage{ae,aecompl}

\usepackage{graphicx}	
\usepackage{amsmath}	
\usepackage{amssymb}	
\usepackage{comment}
\usepackage[dvipsnames]{xcolor}
\usepackage{hyperref}
\hypersetup{colorlinks = true}
\usepackage{multirow}
\usepackage{placeins}
\usepackage{caption}

\newcommand\au{{\,\mathrm{au}}}
\newcommand\yr{{\,\mathrm{yr}}}

\newcommand\msun{{\,\mathrm{M_\odot}}}
\newcommand\rsun{{\,\mathrm{R_\odot}}}
\newcommand\mjup{{\,\mathrm{M_\mathrm{J}}}}
\newcommand\rjup{{\,\mathrm{R_\mathrm{J}}}}
\newcommand\myr{{\,\mathrm{Myr}}}
\newcommand\s{{\,\mathrm{s}}}

\newcommand{\ve}[1]{\boldsymbol{#1}}
\newcommand{\unit}[1]{\hat{\boldsymbol{#1}}}

\newcommand{\moon}{\mathrm{moon}}
\newcommand{\planet}{\mathrm{planet}}
\newcommand{\prim}{\mathrm{prim}}
\renewcommand{\sec}{\mathrm{sec}}

\newcommand{\moonplanet}{{1}}
\newcommand{\planetprim}{{2}}
\newcommand{\primsec}{{3}}
\newcommand{\pert}{\mathrm{pert}}

\newcommand{\irel}{i_{\planetprim\primsec}}
\newcommand{\init}{\mathrm{i}}

\newcommand{\revision}[1]{{#1}}

\title[Can hot Jupiters host exomoons?]{The ominous fate of exomoons around hot Jupiters in the high-eccentricity migration scenario}

\author[A. A. Trani et al.]{
Alessandro A. Trani$^{1,2}$\thanks{E-mail: aatrani@gmail.com}
Adrian S. Hamers,$^{3}$
Aaron Geller,$^{4,5}$ and
Mario Spera$^{4}$
\\
$^{1}$Department of Earth Science and Astronomy, College of Arts and Sciences, The University of Tokyo, 3-8-1 Komaba, Meguro-ku,\\ Tokyo 153-8902, Japan \\
$^{2}$Department of Astronomy, Graduate School of Science, The University of Tokyo, 7-3-1 Hongo, Bunkyo-ku, Tokyo, 113-0033, Japan\\
$^{3}$Max-Planck-Institut f\"{u}r Astrophysik, Karl-Schwarzschild-Str. 1, 85741 Garching, Germany\\
$^{4}$Center for Interdisciplinary Exploration and Research in Astrophysics (CIERA) and Department of Physics and Astronomy,\\ Northwestern University, 1800 Sherman Ave., Evanston, IL 60201, USA\\
$^{5}$Adler Planetarium, Department of Astronomy, 1300 S. Lake Shore Drive, Chicago, IL 60605, USA\\
}

\date{Accepted XXX. Received YYY; in original form ZZZ}

\pubyear{2020}

\begin{document}
\label{firstpage}
\pagerange{\pageref{firstpage}--\pageref{lastpage}}
\maketitle

\begin{abstract}
All the giant planets in the solar system host a large number of natural satellites. Moons in extrasolar systems are difficult to detect, but a Neptune-sized exomoon candidate has been recently found around a Jupiter-sized planet in the {\it Kepler-1625b} system. 
Due to their relative ease of detection, hot Jupiters (HJs), which reside in close orbits around their host stars with a period of a few days, may be very good candidates to search for exomoons. 
It is still unknown whether the HJ population can host (or may have hosted) exomoons. One suggested formation channel for HJs is high-eccentricity migration induced by a stellar binary companion combined with tidal dissipation. Here, we investigate under which circumstances an exomoon can prevent or allow high-eccentricity migration of a HJ, and in the latter case, if the exomoon can survive the migration process. We use both semianalytic arguments, as well as direct $N$-body simulations including tidal interactions. Our results show that massive exomoons are efficient at preventing high-eccentricity migration. If an exomoon does instead allow for planetary migration, it is unlikely that the HJ formed can host exomoons since the moon will either spiral onto the planet or escape from it during the migration process. A few escaped exomoons can become stable planets after the Jupiter has migrated, or by tidally migrating themselves. The majority of the exomoons end up being ejected from the system or colliding with the primary star and the host planet. Such collisions might nonetheless leave observable features, such as a debris disc around the primary star or exorings around the close-in giant.
\end{abstract}

\begin{keywords}
planets and satellites: dynamical evolution
and stability -- planet-star interactions -- binaries: general -- celestial mechanics
 \end{keywords}



\section{Introduction}
The abundance of moons in the Solar system suggests that moons might be common in extrasolar systems. Exomoons might be detected by a number of techniques, including their effect on the transit signal of the host planet (both in transit timing, and duration), or a direct transit signature for large exomoons (see, e.g., \citealt{2018haex.bookE..35H} for an overview). However, despite much effort, no exomoons have been confirmed to date (e.g., \citealt{2012ApJ...750..115K,2013ApJ...770..101K,2013ApJ...777..134K,2014ApJ...784...28K}). \revision{A total of 7 exomoon candidates, including {\it Kepler-1625b I}, have been reported so far \citep{2018SciA....4.1784T,fox2020}; however, the exomoon interpretation of such systems has been put into doubt and is still subject to debate} \citep{2019A&A...624A..95H,2019ApJ...877L..15K,2019arXiv190411896T,2020arXiv200803613K}. Nevertheless, {\it Kepler-1625b I} has opened up questions as to how such massive exomoons could be formed (e.g., \citealt{2018A&A...610A..39H,2018ApJ...869L..27H}). 

The apparent absence of exomoons in detections so far suggests that there is a shortage of satellites around planets---at least, within the range of exoplanets detected to date. The depletion of exomoons has been studied theoretically by a number of authors in a variety of contexts, including migration due to tides \citep{2002ApJ...575.1087B,2016MNRAS.462.2527A,2019MNRAS.489.2313S,2020MNRAS.492.3499S} or protoplanetary disc torques \citep{nam10,2016ApJ...817...18S}, and planet-planet scattering \citep{2007AJ....133.1962N,2013ApJ...769L..14G,2018ApJ...852...85H}. 

Another possibility for exomoons to become unbound from their host planet is excitation of the planet's orbital eccentricity around the parent star by von~Zeipel-Lidov-Kozai (ZLK) oscillations (\citealt{zeipel1910,1962P&SS....9..719L,1962AJ.....67..591K,ito2019}; see \citealt{2016ARA&A..54..441N} and \citealt{shev2017}  for a review and a book) induced by a stellar binary companion. These oscillations, combined with tidal evolution, can shrink the host planet's orbit and transform the planet into a hot Jupiter (HJ; see, e.g., \citealt{2007ApJ...669.1298F,wuyanqin07,naoz2011,2012ApJ...754L..36N,2016MNRAS.456.3671A,2017ApJ...835L..24H,2018AJ....156..128S}). 

In this paper, we study in more detail the latter scenario, and focus specifically on the survivability of exomoons around Jupiter-like planets that are migrating due to the ZLK mechanism with tidal friction. Recently, \citet{2019MNRAS.489.5119M} presented a similar work studying tidal detachment of exomoons around exoplanets excited to high eccentricity through ZLK oscillations induced by a stellar binary companion. Our work can be considered to be complementary, in the sense that we consider in more detail compared to \citet{2019MNRAS.489.5119M} the secular four-body effects. In particular, we take into account the fact that a massive exomoon, through its precession induced on the planetary orbit, can also affect the secular evolution of the planet and even prevent ZLK-driven high eccentricity of the planet in the first place. In addition, we carry out direct four-body integrations of the entire evolution of the system (with the planet starting with a small eccentricity), unlike \citet{2019MNRAS.489.5119M}, who in their four-body integrations focus on the detachment phase when the exoplanet is already highly eccentric. Furthermore, unlike \citet{2019MNRAS.489.5119M}, in our four-body integrations we include tidal interactions between all bodies.

Here we consider moons more massive than solar system counterparts ($m_\moon > 10^{-4} \mjup$). Albeit the mass of natural satellites is considered to be restricted within $10^{-4} m_\planet$, where $m_\planet$ is the host planet mass \citep{canup06}, recent works point out the possibility of large single-moon systems forming in proto-planetary discs \citep{cili18,fujiyu20,moraes20}. Moreover, the exoomon candidate {\it Kepler-1625b I} appears to be in plain violation of the mass scaling relation of satellites in the solary system \citep{hell18}, and massive moons are more likely to be detected with current observational facilities \citep{2020MNRAS.492.3499S}.

The plan of this paper is as follows. In Section~\ref{sec:an}, we estimate the role of exomoons in the high-eccentricity migration process of HJs using analytic arguments. In Section~\ref{sec:nbody}, we confirm and refine our analysis by means of direct $N$-body simulations. In Section~\ref{sec:dis} we discuss our results and extend them to other high-eccentricity migration mechanisms. and conclude in Section~\ref{sec:con}.

\section{Expectations based on semi-analytic arguments}\label{sec:an}

Before presenting detailed $N$-body simulations in Section~\ref{sec:nbody}, we first discuss our expectations of the evolution of exomoons around migrating Jupiter-like planets using semianalytic arguments. Consider a moon (mass $m_\moon$) around a Jupiter-like planet (mass $m_\planet$) in an orbit with semimajor axis $a_\moonplanet$; we will refer to the latter orbit simply as the `lunar orbit'. 
The planet-moon system is orbiting around a primary star (mass $m_\prim$) with semimajor axis $a_\planetprim \gg a_\moonplanet$, and we refer to the latter orbit as the `planetary orbit'. The primary star has a companion star, the secondary star (mass $m_\sec$), in an orbit (the `stellar orbit') with semimajor axis $a_\primsec \gg a_\planetprim$.

If the planetary and stellar orbits are mutually highly inclined (with an inclination $\irel$ close to $90^\circ$), then high-eccentricity ZLK oscillations can be induced in the planetary orbit with a maximum eccentricity approximately given by
\begin{align}
\label{eq:emaxcan}
e_\mathrm{2,max} = \sqrt{1- \frac{5}{3} \cos^2(\irel)}.
\end{align}
Equation~\ref{eq:emaxcan} ignores the presence of the moon and assumes the test particle limit (the planet being much less massive than the stars), the quadrupole-order expansion order only, and zero initial planetary eccentricity. The presence of short-range forces (SRFs) in the planetary orbit (for example, due to general relativity, tidal bulges, and/or rotation) will typically reduce the maximum eccentricity implied by Equation~\ref{eq:emaxcan}. The maximum eccentricity in that case can be calculated semianalytically to quadrupole order using conservation of energy and angular momentum (e.g., \citealt{2002ApJ...578..775B,2007ApJ...669.1298F,2015MNRAS.447..747L}). 

\begin{figure}
	\includegraphics[width=\columnwidth]{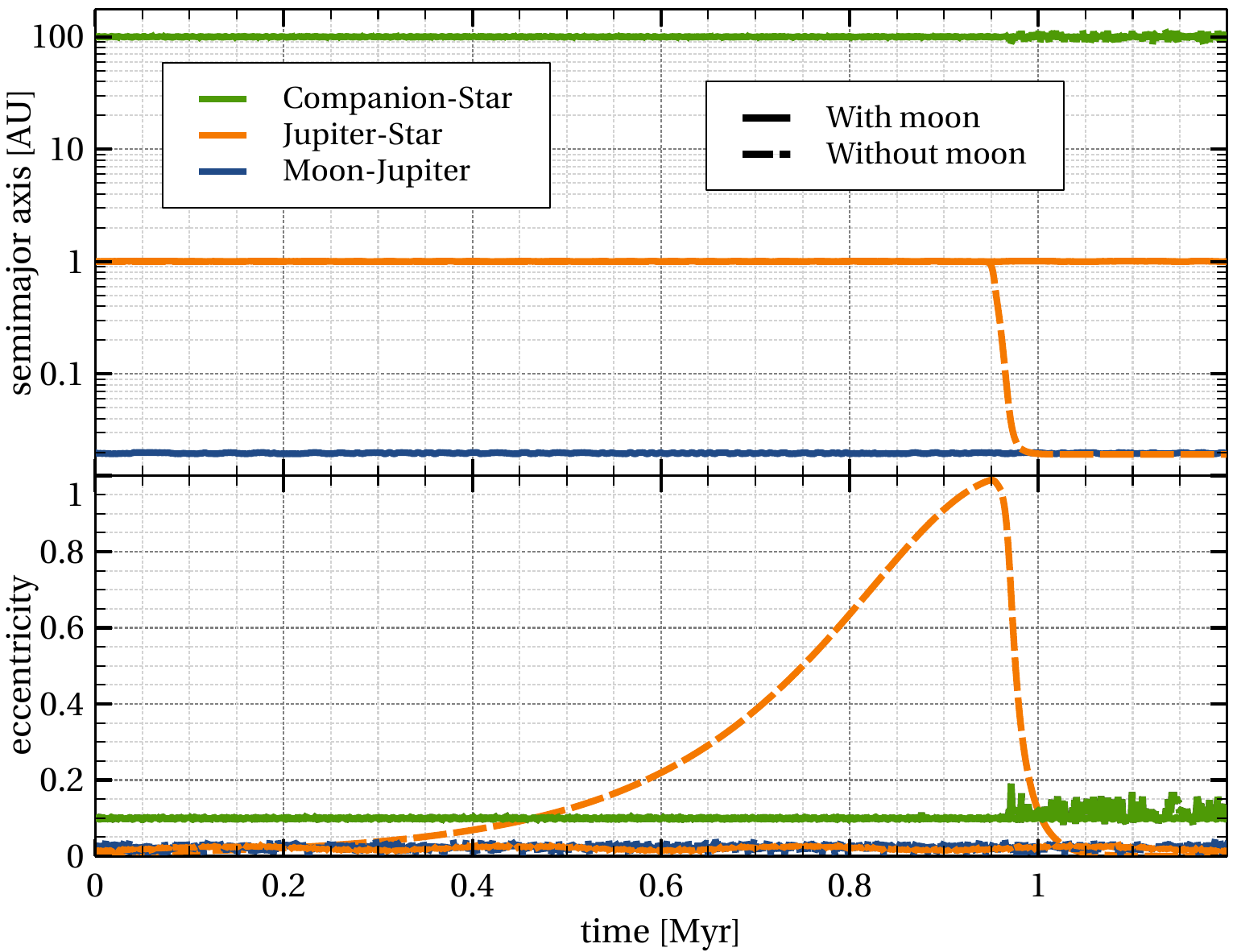}
	\caption{Evolution of semimajor axis and eccentricity of the orbits in two simulations, one with a $0.01 \mjup$ moon orbiting a Jupiter-sized planet, and one without. The companion star has a mass of $0.5 \msun$ and it is inclined by $90^\circ$ with respect to the Jupiter's orbit. The orbits of the moon and the Jupiter are coplanar and prograde. \revision{The moon `shields' the Jupiter from the perturbing companion star, preventing the increase in eccentricity and the tidal migration.}}
	\label{fig:nomoon_vs_moon}
\end{figure}

In an orbit-averaged sense, a moon in orbit around the planet effectively acts as an additional SRF in the planetary orbit.
Therefore, a moon can `shield' the planetary orbit from the secular torque of the stellar companion, and prevent high eccentricities and tidal migration (e.g., \citealt{2015MNRAS.449.4221H,2016MNRAS.455.3180H}). 
An example of the shielding effect (according to an $N$-body integration) is given in Figure~\ref{fig:nomoon_vs_moon}, in which the presence of the moon quenches the excitation of the planetary eccentricity, thus preventing the tidal migration that would otherwise happen in the absence of the moon (see Section~\ref{sec:nbody} for details on the simulation).

Taking into account the secular effects of the moon and restricting to the quadrupole expansion order, the maximum eccentricity in the planetary orbit can be calculated approximately by solving for the algebraic equation for the stationary $e_\planetprim$ based on energy conservation \citep{2015MNRAS.449.4221H}
\begin{align}
\label{eq:emaxmoon}
\nonumber &C_{12} \left(1-e_{\planetprim,\init}^2 \right )^{-3/2} f_\mathrm{q}(\ve{e}_{\moonplanet,\init}, \ve{e}_{\planetprim,\init}, \ve{\jmath}_{\moonplanet,\init},\ve{\jmath}_{\planetprim,\init}) \\
\nonumber &\qquad + C_{23}  \left(1-e_{\primsec,\init}^2 \right )^{-3/2} f_\mathrm{q}(\ve{e}_{\planetprim,\init}, \ve{e}_{\primsec,\init}, \ve{\jmath}_{\planetprim,\init},\ve{\jmath}_{\primsec,\init}) \\
\nonumber \quad &=  C_{12}  \left(1-e_{\planetprim}^2 \right )^{-3/2} f_\mathrm{q}(\ve{e}_{\moonplanet,\init}, \ve{e}_{\planetprim}, \ve{\jmath}_{\moonplanet,\init},\ve{\jmath}_{\planetprim}) \\
&\qquad + C_{23} \left(1-e_{\primsec,\init}^2 \right )^{-3/2}  f_\mathrm{q}(\ve{e}_{\planetprim}, \ve{e}_{\primsec,\init}, \ve{\jmath}_{\planetprim},\ve{\jmath}_{\primsec,\init}),
\end{align}
where we defined the function 
\begin{align}
\nonumber f_\mathrm{q}( \ve{e}_\mathrm{in}, \ve{e}_\mathrm{out}, \ve{\jmath}_\mathrm{in}, \ve{\jmath}_\mathrm{out} ) &\equiv 1 - e_\mathrm{in}^2 + 15 e_\mathrm{in}^2 \left ( \unit{e}_\mathrm{in} \cdot \unit{e}_\mathrm{out} \right )^2 \\
&\quad - 3 \left (1-e_\mathrm{in}^2 \right ) \left ( \unit{\jmath}_\mathrm{in} \cdot \unit{\jmath}_\mathrm{out} \right )^2,
\end{align}
and the constants are
\begin{subequations}
\begin{align}
C_{12} &= \frac{1}{8} \frac{G a_\moonplanet^2}{a_\planetprim^3} \frac{m_\moon m_\planet m_\prim}{m_\moon+m_\planet}; \\
C_{23} &= \frac{1}{8} \frac{G a_\planetprim^2}{a_\primsec^3} \frac{(m_\moon + m_\planet) m_\prim m_\sec}{m_\moon+m_\planet+m_\prim}.
\end{align}
\end{subequations}
The eccentricity and normalized angular-momentum vectors of orbit $k$ are $\ve{e}_k$ and $\ve{\jmath}_k$, respectively; the subscript $\mathrm{\init}$ denotes the initial vector. Stationary points in eccentricity (i.e., minima or maxima) correspond to
\begin{align}
\left (\unit{e}_\planetprim \cdot \unit{e}_{\primsec,\init}\right )^2 = 1 - \left ( \unit{\jmath}_\planetprim \cdot \unit{\jmath}_{\primsec,\init} \right )^2,
\end{align}
whereas the value of $\unit{\jmath}_\planetprim \cdot \unit{\jmath}_\primsec$ at any point (including the stationary point) can be obtained from angular-momentum conservation (neglecting the angular momentum of the lunar orbit), i.e.,
\begin{align}
\nonumber \unit{\jmath}_\planetprim \cdot \unit{\jmath}_{\primsec,\init} &= \frac{1}{2 \sqrt{1-e_\planetprim^2}\sqrt{1-e_{\primsec,\init}^2}} \left [ 2 \sqrt{1-e_{\planetprim,\init}^2} \sqrt{1-e_{\primsec,\init}^2} \, \unit{\jmath}_{\planetprim,\init} \cdot \unit{\jmath}_{\primsec,\init} \right. \\
&\quad \left. + \frac{\Lambda_\planetprim}{\Lambda_{\primsec}} \left ( e_\planetprim^2 - e_{\planetprim,\init}^2\right ) \right ].
\end{align}
Here, $\Lambda_k$ is the circular angular momentum of orbit $k$, i.e.,
\begin{subequations}
\begin{align}
\Lambda_\moonplanet &= \frac{G a_\moonplanet m_\moon m_\planet}{m_\moon+m_\planet}; \\
\Lambda_\planetprim &= \frac{G a_\planetprim (m_\moon + m_\planet)m_\prim}{m_\moon+m_\planet+m_\prim}.
\end{align}
\end{subequations}

Equation~\ref{eq:emaxmoon} is approximate in the sense that it is valid to quadrupole expansion order only, and that the state of the lunar orbit at the stationary point of the planetary orbit is set to the initial one, i.e., $\ve{e}_{\moonplanet} = \ve{e}_{\moonplanet,\init}$ in Equation~\ref{eq:emaxmoon}, and similarly for $\ve{\jmath}_\moonplanet$. In practice, this is a reasonable approximation, since we are interested in the stationary point of $e_\planetprim$ and not of $e_\moonplanet$. 

\begin{figure}
\center
\includegraphics[width = 1.1\columnwidth]{{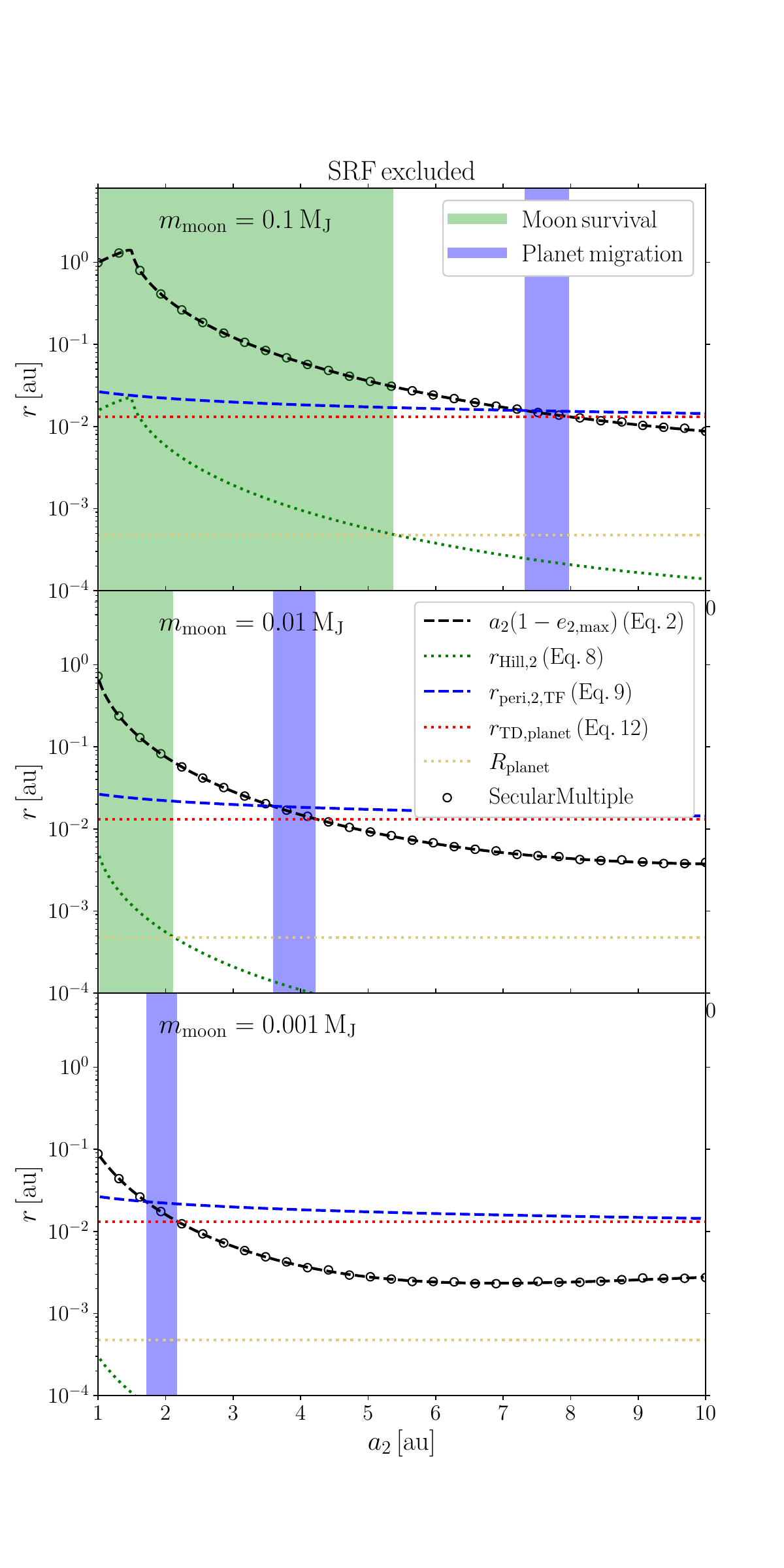}}

\caption[.]{\footnotesize Various distances as a function of $a_\planetprim$. Top, middle and bottom panels correspond to $m_\moon = 0.1, 0.01$ and $0.001 \,\mjup$. Dashed black lines show periapsis distances of the planetary orbit calculated using Equation~\ref{eq:emaxmoon}; black circles show results from numerical solutions of the equations of motion using \textsc{SecularMultiple}. The green dotted line shows the Hill radius of the planet corresponding to Equation~\ref{eq:emaxmoon} (see Equation~\ref{eq:rH}). The blue dashed line shows the critical periapsis distance of the planetary orbit below which we expect the planet to migrate due to tidal dissipation (see Equation~\ref{eq:rperiTF}). The horizontal red dotted line shows the tidal disruption radius of the planet (see Equation~\ref{eq:rTD}). The yellow dotted horizontal lines indicate the planetary radius. The green and blue shaded areas indicate the region where the moon could theoretically survive and the region where the tidal migration is expected to occur, respectively. \revision{Massive moons prevent the formation of a HJ due to the shielding effect, whereas low-mass moons typically cannot survive the high-eccentricity migration process. }}
\label{fig:semianalytic_noSRF}
\end{figure}

In Figure~\ref{fig:semianalytic_noSRF} we show various distances, in particular, the periapsis distance of the planetary orbit, $r_\mathrm{peri,\planetprim} = a_\planetprim(1-e_\planetprim)$, as a function of $a_\planetprim$, whereas other parameters are fixed. We choose three different moon masses: $m_\moon = 0.1$, $0.01$ and $0.001 \,\mjup$ (top to bottom panels). All other parameters are set to $m_\planet=1\,\mjup$, $m_\prim=1\,\msun$, $m_\sec=0.6\,\msun$, $a_\moonplanet= 10^{-3}\,\au$ and $a_\primsec = 600\,\au$. The other (initial) orbital parameters are $e_\moonplanet = e_\planetprim=0.01$, $e_\primsec=0.4$, $i_\moonplanet = i_\planetprim=0.57^\circ$, $i_\primsec=89^\circ$, $\omega_\moonplanet =  \omega_\primsec = 180^\circ$, $\omega_\planetprim = 68.4^\circ$, $\Omega_\moonplanet  = \Omega_\planetprim = \Omega_\primsec = 0.01^\circ$. Here, $i_k$, $\omega_k$, and $\Omega_k$ denote the inclination, argument of periapsis, and longitude of the ascending of orbit $k$, respectively. Note that, with this choice of initial parameters, the initial mutual lunar orbit-planetary orbit inclination is $i_{\moonplanet\planetprim} = 0^\circ$, and the initial mutual planetary-stellar orbit inclination is $\irel = 89^\circ$. 

We compute the maximum $e_\planetprim$ by solving Equation~\ref{eq:emaxmoon}, neglecting other SRFs such as tidal bulges and general relativistic corrections, and show the results in Figure~\ref{fig:semianalytic_noSRF} with black dashed lines. In the absence of the moon, the maximum planetary orbital eccentricity would be instead given by Equation~\ref{eq:emaxcan}. With a moon included, the maximum eccentricity is strongly reduced depending on parameters such as $m_\moon$ and $a_\planetprim$. As $a_\planetprim$ is increased, the `shielding' effect of the moon decreases, and the periapsis distance shrinks.
The circles in Figure~\ref{fig:semianalytic_noSRF} show the periapsis distances obtained by numerically solving the secular equations of motion using \textsc{SecularMultiple} \citep{2016MNRAS.459.2827H,2018MNRAS.476.4139H,2020MNRAS.494.5492H}, and are in good agreement with the semianalytic solutions of Equation~\ref{eq:emaxmoon}.

As the planetary orbit is excited in its eccentricity, its decreased periapsis distance implies that satellites orbiting around the planet could become unbound. Approximately, the orbital radius around the planet for which satellites can remain stable is described by the following ad hoc expression of the Hill radius,
\begin{align}
\label{eq:rH}
r_{\mathrm{Hill},2} = \frac{1}{2} \,a_\planetprim(1-e_\planetprim) \left ( \frac{m_\planet}{3 m_\prim} \right )^{1/3},
\end{align}
where the maximum eccentricity $e_\planetprim$ is obtained from Equation~\ref{eq:emaxmoon}, and which is shown in Figure~\ref{fig:semianalytic_noSRF} with the dotted green lines. The moon is expected to remain bound to the planet as long as its orbital distance around the planet is $\lesssim r_{\mathrm{H},\planet}$. 

Tidal migration of the planet becomes possible only if its eccentricity becomes sufficiently high. In Figure~\ref{fig:semianalytic_noSRF}, we show with the blue dashed lines the periapsis distances of the planetary orbit, $r_{\mathrm{peri},\planetprim,\mathrm{TF}}$, below which we expect tidal dissipation to be efficient to shrink the planetary orbit, and produce a HJ. We estimate the latter boundary by equating the timescale for tidal friction to shrink the orbital semimajor axis by the order of itself (in the limit of $e_\planetprim\rightarrow1$) to the ZLK timescale of the planetary orbit excited by the stellar binary companion, which yields
\begin{align}
\label{eq:rperiTF}
r_{\mathrm{peri},\planetprim,\mathrm{TF}} \sim \frac{a_\planetprim}{2} \left [ \frac{\tau_\mathrm{ZLK}}{\tau_\mathrm{TF}} \beta_a \left (\frac{R_\planet}{a_\planetprim} \right )^8 \right ]^{2/15},
\end{align}
where the ZLK timescale is defined as
\begin{align}
\tau_\mathrm{ZLK} \equiv \frac{P_\primsec}{P_\planetprim} P_\primsec \frac{m_\moon+m_\planet+m_\prim+m_\sec}{m_\sec} \left (1-e_\primsec^2 \right )^{3/2},
\end{align}
and the tidal dissipation-related timescale is
\begin{align}
\tau_\mathrm{TF} \equiv \frac{1}{27} \frac{t_{\mathrm{V,\planet}}}{3 (k_{\mathrm{AM,\planet}} + 2)} \left (\frac{m_\planet}{M_\prim} \right )^2,
\end{align}
with $t_{\mathrm{V,\planet}}$, $k_{\mathrm{AM,\planet}}$, and $R_\planet$ the viscous timescale, apsidal motion constant, and radius of the planet, respectively. Furthermore, $\beta_a \equiv 451/160$, and $P_k$ denotes the orbital period of orbit $k$. Equation~\ref{eq:rperiTF} was derived by assuming the equilibrium tide model \citep{1981A&A....99..126H}, and assuming pseudosynchronisation (i.e., that the spin frequency is close to the orbital frequency at periapsis). Here, we set $k_{\mathrm{AM,\planet}} = 0.19$, $t_\mathrm{V,\planet} = 1.3 \times 10^4\,\mathrm{hr}$ \citep{2012arXiv1209.5724S}. 

The condition $r_{\mathrm{peri},\planetprim} < r_{\mathrm{peri},\planetprim,\mathrm{TF}}$ is not sufficient for successful migration of the planet, since the planet could be tidally disrupted if it ventures too close to the primary star. Specifically, the latter is expected to occur if $r_{\mathrm{peri},\planetprim} < r_\mathrm{TD,\planet}$, where 
\begin{align}
\label{eq:rTD}
r_\mathrm{TD,\planet} = \eta R_\planet \left ( \frac{m_\prim}{m_\planet} \right )^{1/3},
\end{align}
where we adopt $\eta=2.7$ \citep{2011ApJ...732...74G}. Finally, we should consider that, for survival of the moon, the lunar orbit should, evidently, at least be larger than $R_\planet$; in Figure~\ref{fig:semianalytic_noSRF}, we show a yellow dotted horizontal line indicating  $r=R_\planet$. 

Figure~\ref{fig:semianalytic_noSRF}, which does not include additional SRFs, paints the following picture: even a relatively low-mass moon ($10^{-3}\,\mjup$; bottom panel) is able to effectively shield the planet, and prevent excitation of the planetary orbit, unless $a_\planetprim$ is large $(\gtrsim 5\,\au$). \revision{The shielding effect originates from an effective additional SRF acting on the orbit of the planet around the primary star. This SRF is due to the quadrupole moment of the moon, and causes the circulation of the planetary argument of pericenter, suppressing the ZLK oscillations.} However, for large $a_\planetprim$ and low $m_\moon$, the excited planetary eccentricity is very high, and the planet is expected to be tidally disrupted, rather than to tidally migrate. Even if the planet manages to survive migration, the small Hill radius during the migration phase would imply that no moon could survive the process. 

The region in $a_\planetprim$ space that allows for the planetary migration never overlaps with the region in which the moon can remain bound to the planet. This is especially true for low-mass moons; for higher-mass moons the two regions get closer, but still do not overlap, i.e., the planet's Hill radius is so small that no moon could feasibly remain in a stable orbit around it \citep{doming06}.

For higher-mass moons ($0.1\mjup$; top panel), the shielding effect is much more severe, and even planets at $10\,\au$ are still affected by the presence of the moon. The planetary orbit can nevertheless become sufficiently eccentric to potentially tidally migrate if $a_\planetprim \sim 8\,\au$ (although the planet in that case is also close to being tidally disrupted). However, the Hill radius for that $a_\planetprim$ is sufficiently small that moons are not expected to survive the migration process. 

The above picture remains unchanged when also considering other short-range force such as relativistic and tidal precession (see the Appendix and Figure~\ref{fig:semianalytic_SRF}).
On the other hand, these arguments are based on the secular (i.e., orbit-averaged) approximation, and ignore many potentially important effects such as non-secular evolution, and tides in the lunar orbit in addition to those in the planetary orbit. In Sect.~\ref{sec:nbody}, we carry out more detailed $N$-body simulations to address these caveats.

\section{$N$-body simulations}\label{sec:nbody}

\subsection{Numerical setup}
We employ the \texttt{TSUNAMI} code (A.A.Trani, in prep.) to directly integrate the 4-body system consisting of a moon-hosting planet, the parent star and the stellar companion. \texttt{TSUNAMI} integrates the equations of motion derived from a logarithmic Hamiltonian in an extended phase space \citep{mik99a}, using a chain coordinate system to reduce round-off errors \citep{mik90}, combined with Bulirsh-Stoer extrapolation to increase accuracy \citep{sto80}. We include the first-order post-Newtonian correction to the gravitational acceleration, and the tidal interaction force from \citet{1981A&A....99..126H}. 

As with the analytic estimates, we drop the spin-orbit tidal coupling term, i.e. we assume mutual pseudo-synchronization at every timestep. Tidal interactions between all the bodies in the simulations are considered. We set the apsidal motion constant $k_{\mathrm{AM}} = 0.1$ for the planet and the moon, and $k_{\mathrm{AM}} = 0.014$ for the two stars. In the \citet{1981A&A....99..126H} model, the efficiency of the tides is parametrized by the time-lag of the bulges. We set the time-lag to $\tau = 0.66 \times 10^2 \s$ for the planet and moon, and to $\tau = 0.15 \s$ for the two stars, respectively. The time-lag for the planet and the moon is about $10^2$ higher than what is estimated for high-eccentricity migration \citep{2012arXiv1209.5724S}; we make this choice in order to shorten the computational time of the simulations. Such approach has been commonly used in similar studies \citep[e.g.][]{anto2016}. The time-lag and tidal efficiency are nonetheless largely uncertain and strictly depend on the rheology of the bodies \citep{ogil13,maka14,efro14}.
 
The radius of the bodies is chosen to be $1 \rsun$ for the star and $1 \rjup$ for the planet. The radius of the moon is calculated from its mass following \citet{chen_kipping17}, while the radius of the companion in solar radii is given by $(m_\sec/1\msun)^{0.881}$ \citep{kippenhahn12}.
We also check for collisions at every timestep; if a collision between two bodies is detected the integration is automatically stopped. 

\subsection{Comparison with our analytic estimate}\label{sec:numgrid}

\begin{table}
	\centering
	\caption{Initial parameters for the controlled grid of simulations.}
	\label{tab:ic_contro}
	\begin{tabular}{lc} 
		\hline
		Parameter & Value \\
		\hline
		$m_\prim$ $\quad[\msun]$ & $1$ \\
		$m_\sec$ $\quad[\msun]$ & $0.6$ \\
		$m_\planet$ $\quad[\mjup]$ & $1$ \\
		$m_\moon$ $\quad[\mjup]$ & $0.1$, $0.01 \mjup$ \\
		$a_\primsec$ $\quad[\au]$ & $600$ \\
		$a_\planetprim$ $\quad[\au]$ & $1$--$10$ \\
		$a_\moonplanet$ $\quad[\au]$ & $0.01$\\
		$e_\primsec$ & $0.4$ \\
		$e_\planetprim$ & $0.01$ \\
		$e_\moonplanet$ & $0.01$\\
		$i_{12}$ & $0$\\
		$i_{23}$ & $\pi/2$ \\
		$\omega_\primsec$ & $\pi$\\
		$\omega_\planetprim$ & $0.38\pi$ \\
		$\omega_\moonplanet$ & $\pi$ \\
		$\Omega_\primsec$ & $0.01$ \\
		$\Omega_\planetprim$ & $0.01$ \\
		$\Omega_\moonplanet$ & $0.01$ \\
		$M_\primsec$ & $\pi$ \\
		$M_\planetprim$ & $0.3\pi$ \\
		$M_\moonplanet$ & $1.66\pi$ \\
		\hline
	\end{tabular}
	\begin{flushleft}
	{\scriptsize 
	$m$: mass; $a$: semimajor axis; $e$: eccentricity; $i$: inclination; $\omega$: argument of periapsis; $\Omega$: longitude of the ascending node; $M$: mean anomaly. The subscripts $_1$, $_2$ and $_3$ refer to the primary-secondary orbit, the planet-primary orbit and the moon-planet orbit, respectively. All angles are expressed in radians.}
	\end{flushleft}
\end{table}

We first run a controlled grid of simulations to better compare our analytic estimate with the results from direct integration. We fix all the initial parameters but the semimajor axis of the planet's orbit $a_\planetprim$, which ranges between 1 and $10\au$. All the initial parameters are listed in Table~\ref{tab:ic_contro}. We run two sets of simulations for different lunar masses: $m_\moon = 0.01 \mjup$ and $m_\moon = 0.1 \mjup$. Each set consists of 200 realizations with different $a_\planetprim$. The simulations were run for $2 \times 10^7 \yr$. Table~\ref{tab:controut} summarizes the outcomes of the simulations.

\begin{figure}
	\includegraphics[width=\columnwidth]{{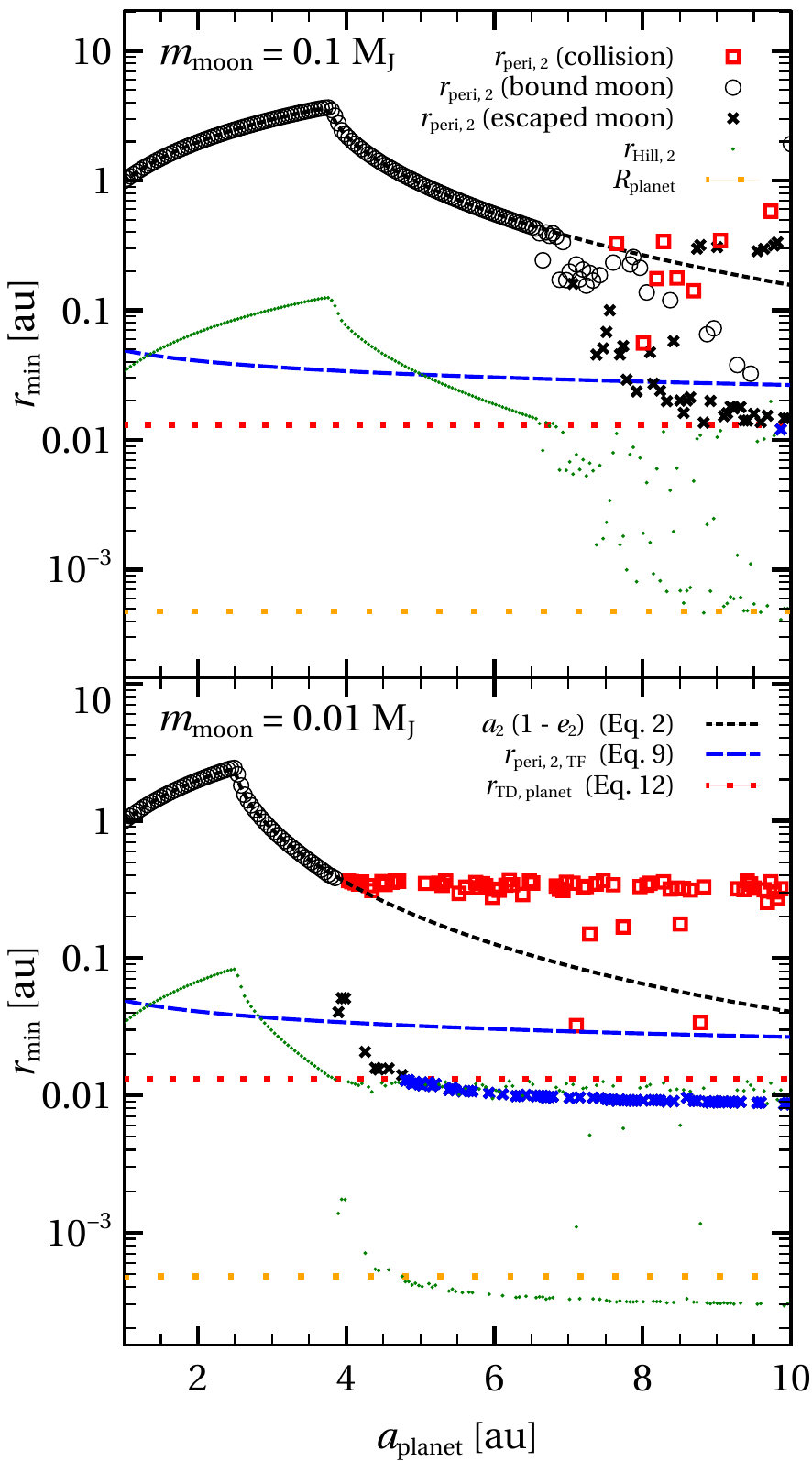}}
	\caption{\revision{Same as Figure~\ref{fig:semianalytic_noSRF}, but compared to the direct N-body simulations.} Various distances as a function of the initial semimajor axis of the planet. Each marker is obtained from a single simulation, lines are obtained from analytic arguments as in Section~\ref{sec:an}. Empty circle, cross or square show the minimum periapsis distance of the planet. The marker shape denotes the end-state of the system. Empty circle: the moon is still bound to the planet. Cross (any color): the moon escaped from the planet. Red square: the simulation stopped because a collision occurred between two bodies. Black (blue) marker: the final planetary semimajor axis is larger (smaller) than $1\au$. Green dot: Hill radius corresponding to the minimum periapsis distance.
		Dashed black line: minimum periapsis distance of the planet as obtained from Equation~\ref{eq:emaxmoon}. Dot-dashed black line: minimum periapsis distance allowed by the quadrupole approximation alone, neglecting the moon (Equation~\ref{eq:emaxcan}). Blue dashed line: minimum periapsis distance that allows migration of the planet (Equation~\ref{eq:rperiTF}). Red dotted line: tidal disruption radius of the planet (Equation~\ref{eq:rTD}). Orange dotted line: planetary radius. Top panel: $m_\moon = 0.1 \mjup$; bottom panel: $m_\moon = 0.01 \mjup$. \revision{Note how the results from the simulations start to diverge from the semi-analytical results as soon as the planetary Hill radius becomes comparable to the moon's semimajor axis at $0.01 \au$}. 
	}
	\label{fig:rmin_aplanet}
\end{figure}

Figure~\ref{fig:rmin_aplanet} shows the minimum periapsis distance and other quantities obtained from the simulations, along with the analytic estimates analogue as in Figure~\ref{fig:semianalytic_noSRF}. There is a tight agreement between the minimum periapsis distance expected from Equation~\ref{eq:emaxmoon} and the results from the simulations until $r_\mathrm{Hill,2}\sim a_{1,i}$, i.e. when the planetary Hill radius becomes comparable to the initial semimajor axis of the moon. When this occurs, the moon becomes dynamically unstable and the secular approximation does not hold anymore. As expected, less massive moons become dynamically unstable at smaller $a_2$, because the shielding effect is lower and the Jupiter can reach higher eccentricities at the same $a_2$.

\begin{table}
	\centering
	\caption{Outcome fractions of the controlled grid simulations. Left column: set with low-mass moons. Right column: set with high-mass moons. Each set consists of 200 realizations.}
	\label{tab:controut}
	\begin{tabular}{lcc} 
		\hline\hline
		\multirow{2}{*}{Collisions} & \multicolumn{2}{c}{$m_\moon$} \\
		 &  $0.01 \mjup$ &  $0.1 \mjup$  \\
				\hline	
		Planet-star & - & 0.015 \\
		Moon-star & 0.27 & 0.005 \\
		Moon-planet & 0.06 & 0.02 \\\hline
		Total & 0.33 &  0.04 \\
		\hline\hline
		\multicolumn{3}{l}{Moon escaped}   \\ 
		\hline
		From planet & 0.555 & 0.215 \\ 
		From system & 0.345 & 0.145 \\
		\hline\hline
		\multicolumn{3}{l}{Planet escaped}   \\ 
		\hline
		From system & - & 0.045 \\ 
		\hline\hline
		\multicolumn{3}{l}{Planet migration within $1 \au$}   \\ 
			 & 0.315 & 0.015 \\ 
		\hline\hline
		\multicolumn{3}{l}{Moon-turned-planet migration within $1 \au$}   \\ 
		& - & 0.055 \\ 
		\hline\hline
		\multicolumn{3}{l}{Planet+moon migration within $1 \au$}   \\ 
		& - & - \\ 
		\hline\hline
	\end{tabular}
	\begin{flushleft}
	\end{flushleft}
\end{table}

\subsubsection{Dynamical instability of low-mass moons}

\begin{figure*}
	\includegraphics[width=0.477\linewidth]{{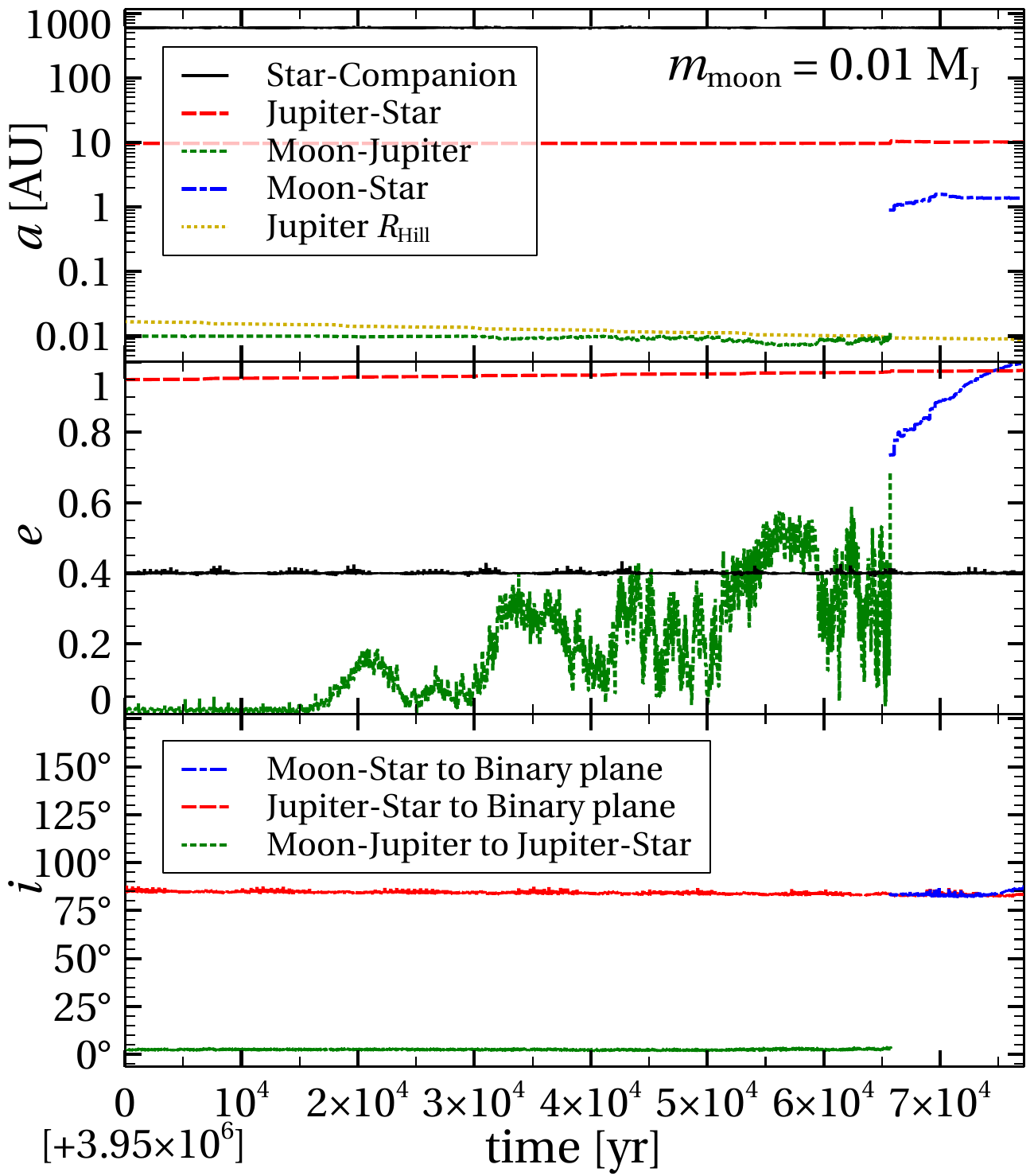}}
	\includegraphics[width=0.477\linewidth]{{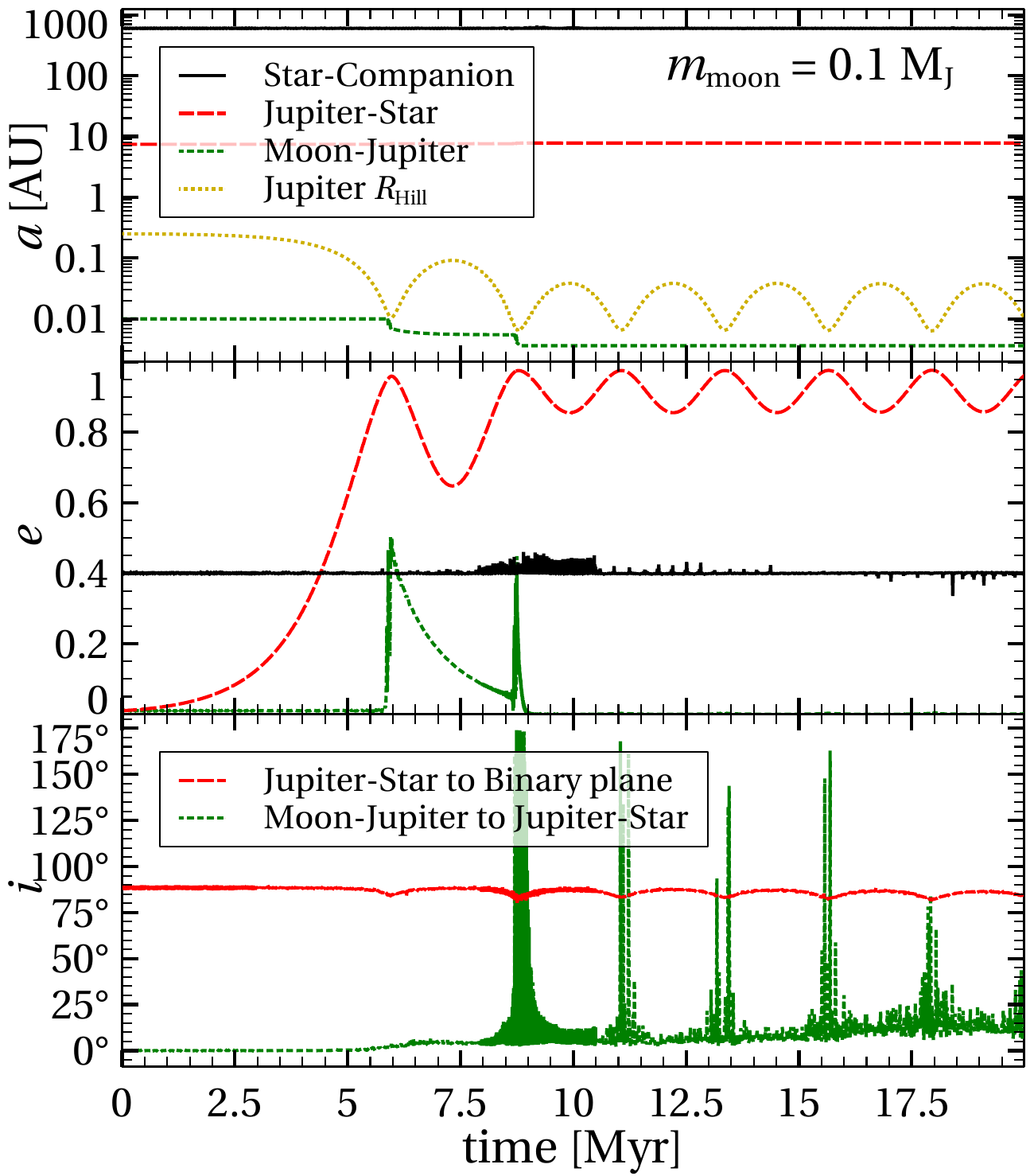}}
	\includegraphics[width=0.477\linewidth]{{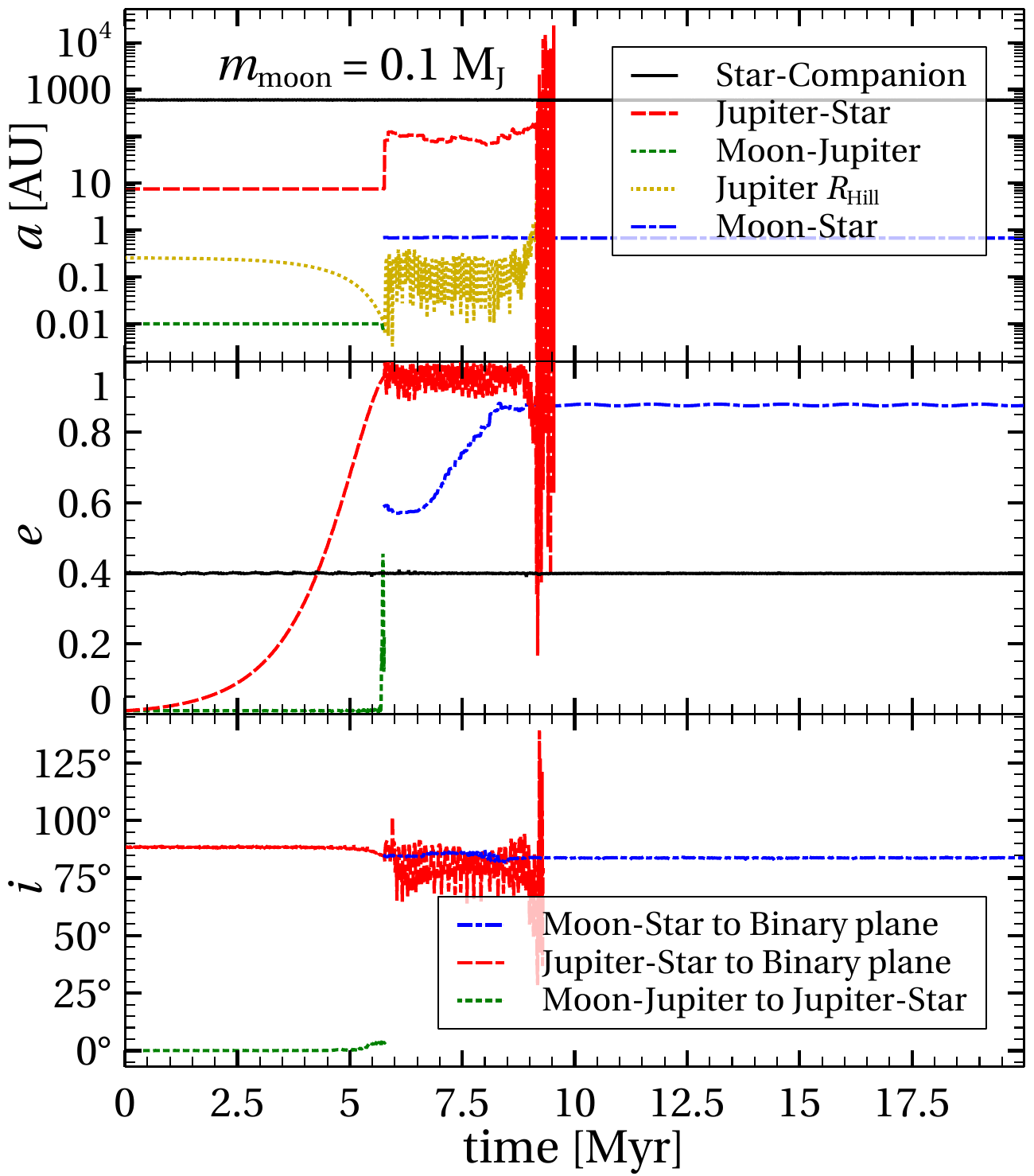}}
	\includegraphics[width=0.477\linewidth]{{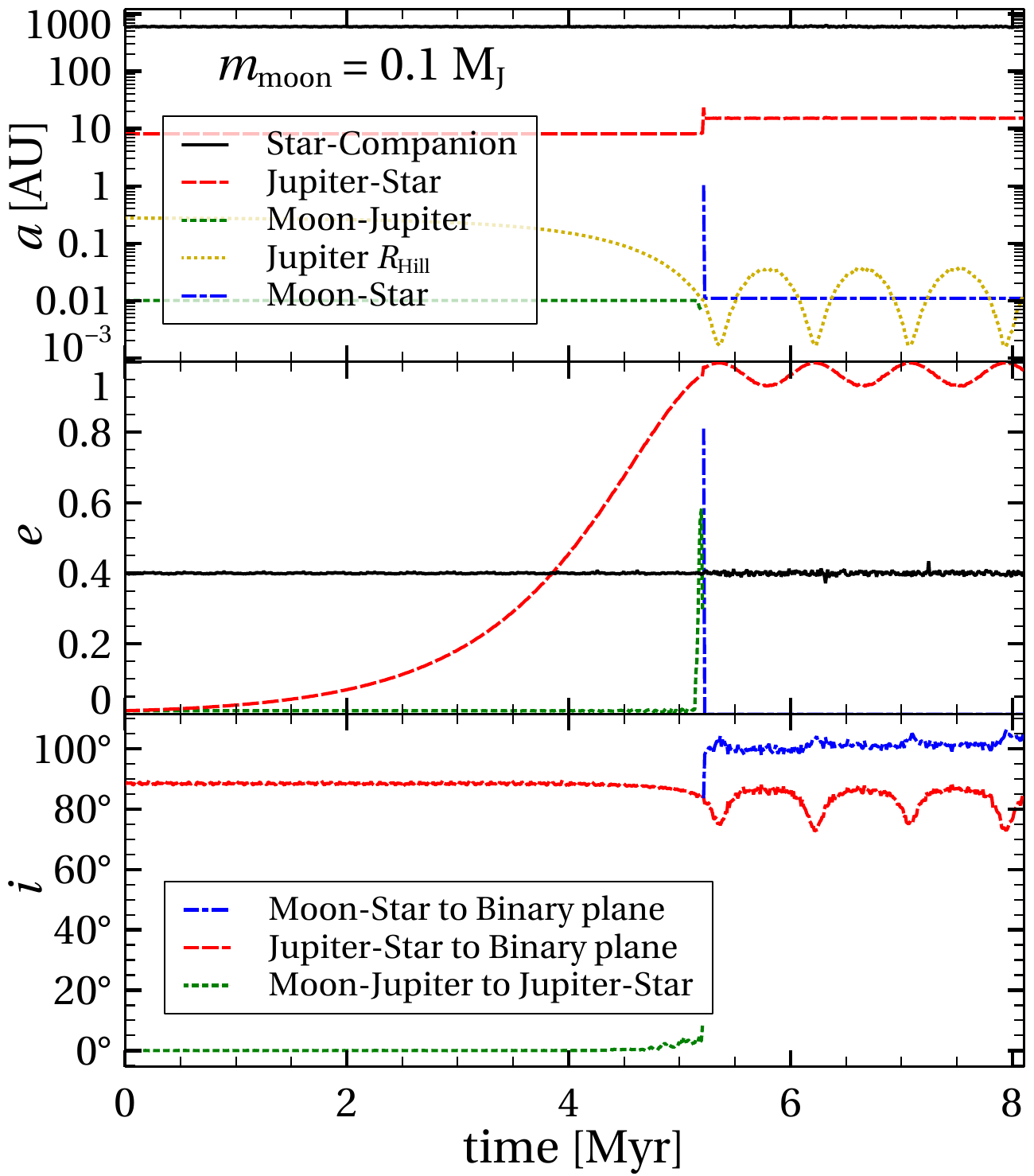}}
	\caption[.]{\footnotesize Evolution of orbital parameter for 4 different realization of the simulation grid of Section~\ref{sec:numgrid}. Each realization represents a possible outcome of the system evolution after the moon becomes dynamically unstable. Top panels: semimajor axis. Middle panels: eccentricity. Bottom panels: mutual inclination. Top-left panel: collision of the moon with the primary star following lunar escape, which is the most common outcome for low-mass moons. Top-right: tidal decay of the moon following the excitation of its eccentricity due to the dynamical instability. Bottom-left panel: three-body scattering of the planet-moon system with the primary star, with inward scattering of the moon and outward scattering of the planet. The planetary orbit gets subsequently ejected due to dynamical instability triggered by the secondary star. Bottom-right panel: tidal migration of the moon around primary star after the three-body scattering. }
	\label{fig:outcome_example}
\end{figure*}

For low-mass moons, ($m_\moon = 0.01 \mjup$), the most common outcome after the dynamical instability is the escape of the moon from the planet. When this occurs, in about $60\%$ of the simulations the moon gets immediately ejected from the system during the periapsis passage of the planet. In the other $40\%$, the moon becomes a planet orbiting the primary star. This phase is only temporary, however: once the moon escapes, its orbit is strongly perturbed by the planet and its eccentricity grows until a collision with the star occurs. This outcome is shown in the top-left panel of Figure~\ref{fig:outcome_example}. In 12 realizations, the moon collides with the planet instead. In only one realization the moon becomes a planet, because the planet dynamically decouples from the moon by tidal circularization before the moon can collide with the star.

After the moon escapes, the planetary eccentricity can be freely excited by the ZLK mechanism and the Jupiter can tidally migrate. In $63$ realizations the final planetary semimajor axis becomes less than $1 \au$ due to tidal circularization (blue crosses in Figure~\ref{fig:rmin_aplanet}). Notice, however, that in all migration cases, the minimum periapsis distance during the migration is less than the tidal disruption radius (red dotted line in Figure~\ref{fig:rmin_aplanet}), suggesting that the Jupiter will be tidally disrupted in the process. 

\subsubsection{Dynamical instability of massive moons}

For higher moon masses, the dynamical instability of the moon triggers an interesting interplay between tidal dissipation and the ZLK mechanism. Once $r_\mathrm{Hill,2}\sim a_{1,i}$,  the eccentricity of the lunar orbit is excited, which triggers the tidal circularization of the lunar orbit. After the lunar orbit becomes tighter, the shielding effect is smaller and the eccentricity of planet can be excited further, so that the minimum periapsis distance is smaller than the one expected from the analytic estimate, which assumes energy conservation (i.e. no change in semimajor axis). In the top-right panel of Figure~\ref{fig:outcome_example} we show the typical evolution of the orbital parameters in this scenario. Such systems can be thus identified in the bottom panel of Figure~\ref{fig:rmin_aplanet} as the empty circles below the dashed black line that denotes the semi-analytic result from Equation~\ref{eq:emaxmoon}. Only in a few simulations, the tidal decay causes the inspiral and collision of the moon onto the planet.

Another unforeseen outcome is the ejection of the planet from the system. In many cases, the planet-moon system effectively undergoes a three-body encounter with the primary star. This often leads to the prompt ejection of the moon and, in 9 simulations, even the ejection of the planet. 

The planet can also be scattered on an outward orbit, while the moon remains on an inner orbit around the primary star. However, the new orbit of the planet might be unstable due to the secondary star: in this case, it will be ejected from the system following a scattering with the secondary. This latter case is shown in the bottom-left panel of Figure~\ref{fig:outcome_example}. 
\revision{By this scattering mechanism, 10 moons end up as planets around the primary star on a orbit at less then 1 au. In one case, after becoming a planet, the moon migrates to a short-period orbit via high-eccentricity tidal circularization (bottom-right panel of Figure~\ref{fig:outcome_example}).}

\subsection{Population synthesis study}
To investigate the role of moons in a more general and realistic setup, we first generate a Monte Carlo set of star-planet-companion systems in the following way. The mass of the primary star and the planet are fixed to $1 \msun$ and $1 \mjup$, respectively, while the secondary mass is drawn from a uniform distribution between $0.08$ and $0.6$ \citep{ngo16}. The binary orbital period and eccentricity are drawn from \citet{ragh10}. The semimajor axis of the planet is uniformly sampled between 1 and $5 \au$. We also impose that the planet fulfills the stability criterion of \citet{hol1999}.
The inclination $i_{23}$ between the orbit of the companion and the orbit of the planet is drawn uniformly in $\cos{i_{23}}$ between $70^\circ$ and $110^\circ$.

We generate a total of 6000 triple systems, and we evolve them for 2 Myr. At the end of the run, we select those simulations in which the Jupiter migrates below $1 \au$ and re-run them adding a moon around the planet. In total we find 1854 simulations in which the Jupiter migrates below $1 \au$ without colliding with the star. \revision{Afterwards, we randomly select 1000 initial conditions from these 1854 runs, and add a moon around the Jupiter.} The semimajor axis of the moon is uniformly sampled between $2(R_\planet + R_\moon)$ and $r_{\mathrm{Hill},2}$, and its eccentricity is set to $0.01$. We perform a total of 2 sets of 1000 realizations, one with $m_\moon = 0.1 \mjup$ and one with $m_\moon = 0.01 \mjup$, and run them for $10 \myr$.

\begin{table}
	\caption{Outcome fractions of the Monte Carlo simulations. Left column: set with low-mass moons. Right column: set with high-mass moons.}
	\label{tab:sampout}
	\begin{tabular}{lll} 
		\hline\hline
		\multirow{2}{*}{Moon fate} & \multicolumn{2}{c}{$m_\moon$} \\
		&  $0.01 \mjup$ &  $0.1 \mjup$  \\
		\hline
		Bound to planet (efficient shielding) & 0.006 & 0.091 \\ 
		Inspiral on planet & 0.186 & 0.196 \\
		Collision with primary at planet periapsis & 0.296 & 0.227 \\
		Escape from planet & 0.512  & 0.486 \vspace{1.5pt} \\ 
	    \multicolumn{3}{l}{Post-escape fate:} \\
		\quad collision with primary & 0.138 & 0.122 \\
		\quad collision with planet & 0.003 & 0.009 \\
		\quad turned primary planet & 0.013 & 0.064 \\ 
		\quad turned circumbinary planet & - & 0.001 \\
		\quad ejected from system & 0.358 & 0.290 \\ 
							
		\hline\hline
		\multirow{2}{*}{Collisions} & \multicolumn{2}{c}{$m_\moon$} \\
		\hline	
		Planet-primary & 0.024 & 0.037 \\
		Moon-primary & 0.434 & 0.349 \\
		Moon-planet & 0.189 & 0.205 \\\hline
		Total & 0.647 &  0.591 \\
		\hline\hline
		
		\multicolumn{3}{l}{Planet migration within $1 \au$}   \\ 
		& 0.172 & 0.150 \\ 
		\hline\hline
		\multicolumn{3}{l}{Moon-turned-planet migration within $1 \au$}   \\ 
		& - & 0.022 \\ 
		\hline\hline
		\multicolumn{3}{l}{Planet+moon migration within $1 \au$}   \\ 
		& - & - \\ 
		\hline\hline
	\end{tabular}
	\begin{flushleft}
	\end{flushleft}
\end{table}

Table~\ref{tab:sampout} summarizes the outcome of the Monte Carlo simulations. The results confirm the trends presented in Section~\ref{sec:numgrid}: there are no cases of planets successfully migrating together with the moon. 

Small moons are less likely to remain bound to their host planets, and more likely to collide with the primary star. More massive moons induce more efficient shielding, so that in about $9\%$ of the systems the planet remains on its original orbit along with its moon. 

In all the other cases, the moon becomes dynamically unstable, and either in-spirals on the planet, collides with the star or escapes from the planet. The moon can be ejected immediately from the system or remain orbiting the primary star as a planet. 

This latter state is however temporary, because once the moon escapes, its orbit remains very close to that of the host planet \citep[e.g.][]{tra16b}. Thus, the moon undergoes scatterings with the planet until there is a collision, an ejection or the two orbits dynamically decouple via tidal circularization around the primary. 

\begin{figure}
	\includegraphics[width=1\linewidth]{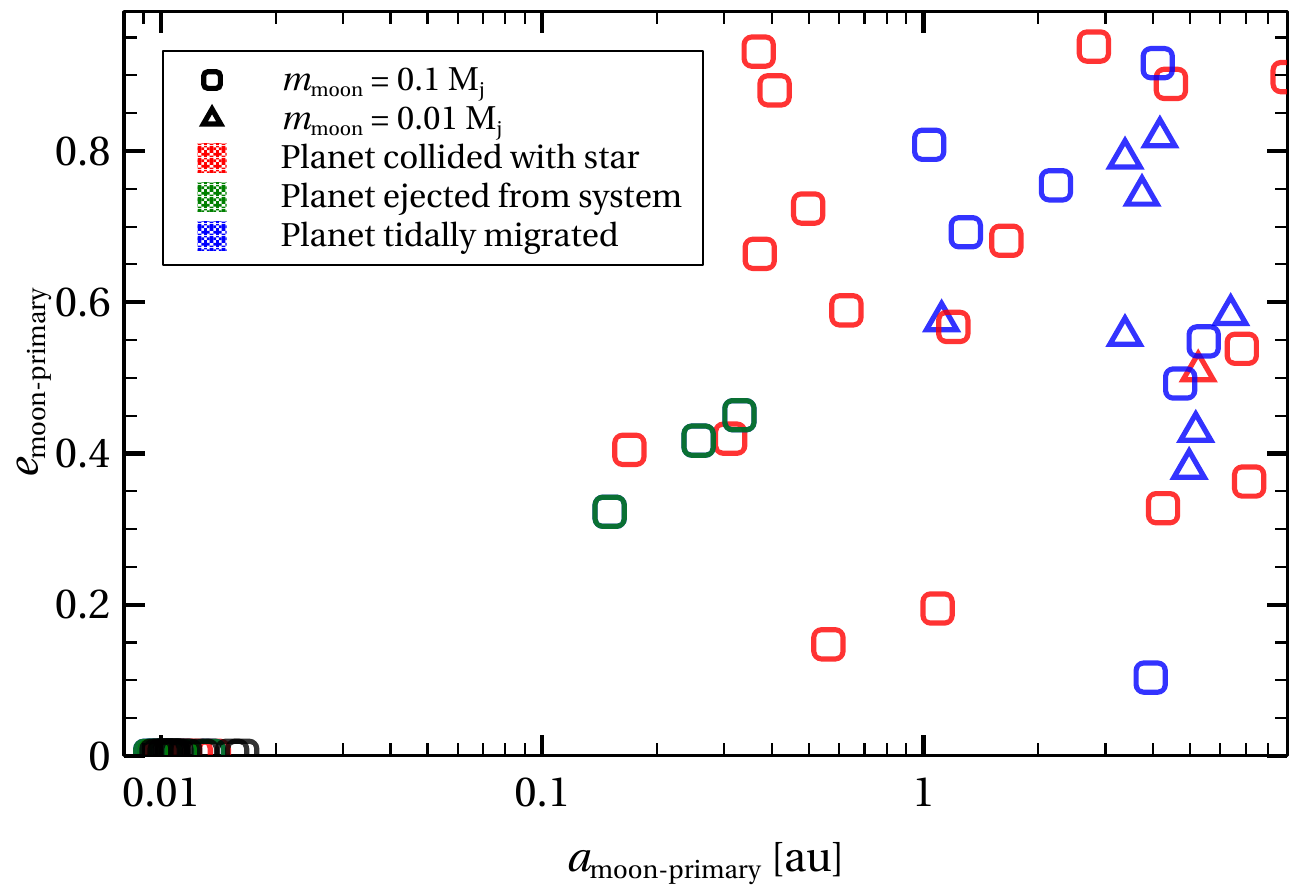}
	\caption[.]{\footnotesize Semimajor axis and eccentricity of stable moon-turned-planets around the primary after escaping the host planet. Squares: $0.1 \mjup$ moons. Triangles: $0.01 \mjup$ moons. The color indicates the pathway that led the lunar orbit to become stable. Red color: the host planet collided with the star. Green color: the host planet was ejected from the system. Blue color: the host planet tidally migrated close to the star. \revision{The cluster at ${\sim}0.01 \au$ and zero eccentricity is constituted by moons that have tidally migrated around the primary star.}}
	\label{fig:aemoon_post}
\end{figure}

Figure~\ref{fig:aemoon_post} shows the semimajor axis and eccentricity of stable moons-turned-planets. The orbit of the moon can become stable only if (1) the moon or the planet tidally circularize turning into short-period planets, (2) the planet collides with the star or (3) the planet gets ejected (only possible for massive moons). The color of the markers in Figure~\ref{fig:aemoon_post} indicates the scenario that led the moon to become a stable planet.

About $6\%$ of the $0.1 \mjup$ moons can become a stable planet around the primary. Of those, more than $33\%$ undergo tidal migration around the primary, becoming hot Neptunes. Due to the higher number of collision and ejections from the system, fewer $0.01 \mjup$ moons turn into planets. 
A substantial fraction of moons (${\sim}36\%$ and ${\sim}29\%$ for $0.01$ and $0.1 \mjup$, respectively) get ejected from the system, becoming free-floating planets. There is a very small probability ($0.1\%$) that a $0.1 \mjup$ massive moon can become a circumbinary planet, orbiting the binary in a P-type orbit.

\section{Discussion}\label{sec:dis}

\subsection{Relevance to other high-eccentricity migration models}

Both the $N$-body simulations and the semi-analytic arguments indicate that it is highly unlikely for HJs to migrate inwards while retaining their moons in the binary-ZLK migration model. Moreover, the moon can actually damp the excitation of the planetary eccentricity, preventing the tidal decay of the Jupiter.


It is reasonable to ask to what extent our results hold in the other high-eccentricity migration scenarios, such as the coplanar model \citep{petro2015a}, the secular chaos model \citep{wuyanqin11,2017MNRAS.464..688H}, the ZLK model with a planetary perturber \citep[e.g.][]{naoz2011,ham17}, or the planet-planet scattering model \citep{rasio96,weid96,chat08}. Assessing the survivability of moons in these scenarios would require detailed calculations, but it is possible to estimate the effect of the presence of the moon with simple analytic considerations.

The picture outlined in Section~\ref{sec:an} is \revision{largely} unchanged when the ZLK oscillations are induced by a massive planetary perturber. In general, the shielding effect can be approximately estimated by comparing the quadrupole ZLK timescale $T_\mathrm{ZLK}$ of the inner and outer triple systems, constituted by the moon-planet-star system and the planet-star-perturber system, respectively. When inner ZLK timescale is much shorter than the outer one, $T_\mathrm{ZLK,inn} \ll T_\mathrm{ZLK,out}$, the ZLK oscillations of the outer orbit will be quenched, i.e. \citep{2015MNRAS.449.4221H}:
\begin{align}\label{eq:shield}
& \frac{T_\mathrm{ZLK,inn}}{T_\mathrm{ZLK,out}} = \nonumber \\
& \quad\quad  = \left(\frac{a_\planetprim^3}{a_\moonplanet a_\primsec^2}\right)^{3/2} \left(\frac{m_\moon + m_\planet}{m_\moon + m_\planet + m_\prim}\right)^{1/2} \cdot \nonumber \\ 
 & \quad\quad \cdot \frac{m_\pert}{m_\prim} \left(\frac{1 - e_\planetprim^2}{1 - e_\primsec^2}\right)^{3/2} \ll 1
\end{align}
where $m_\pert$ is the mass of the perturber, whether a companion star or a planet.
The main difference with respect to a stellar perturber is that, to compensate for the smaller mass ($T_\mathrm{ZLK,out} \propto 1/m_\pert$), the perturbing planet needs to be very close to the orbit of the Jupiter, in order for the outer ZLK timescale to be shorter than the inner one.

Our analysis also applies to the coplanar model proposed by \citet{petro2015a}, wherein the planetary eccentricity is excited by an outer planet lying in the same orbital plane. In this case, the same Hamiltonian expansion used for the ZLK mechanism can describe the evolution of the system. However, differently from the mutually inclined case, there is no angular momentum exchange between the two planets at the quadrupole-order approximation. In fact, the eccentricity growth occurs on the timescale of the octupole-order approximation. 
As described in Section~\ref{sec:an}, the presence of the moon acts as an additional short-range force that causes the apsidal precession of the planetary orbit. In this sense, we can estimate the shielding effect by comparing the quadrupole time scale of the moon with the octupole timescale from the perturbing planet. From \citet{antognini2015}, $T^\mathrm{oct}_\mathrm{ZLK} = T^\mathrm{quad}_\mathrm{ZLK} / \sqrt{\epsilon_\mathrm{oct}}$, where 
\begin{equation}
\epsilon_\mathrm{oct} = \frac{a_\planetprim}{a_\primsec} \frac{e_\primsec}{1 - e_\primsec^2}
\end{equation}
Hence, we can write the shielding condition (Equation~\ref{eq:shield}) for the coplanar scenario as 
\begin{equation}
\frac{T_\mathrm{ZLK,inn}}{T_\mathrm{ZLK,out}} \sqrt{\epsilon_\mathrm{oct}} \ll 1
\end{equation}
Since the octupole timescale is longer than the quadruple one, the moon shielding effect is increased in this scenario: moons can better shield perturbations from a coplanar perturber compared to an inclined pertuber. 

It is more challenging to predict the role of moons in the secular chaos model of \citet{wuyanqin11}. In this scenario, the eccentricity growth is due to the overlap of higher order secular resonances, e.g. between apsidal or nodal precession frequencies, in a multi-planet system. The moon presence would affect such frequencies, likely shifting the loci in phase space where eccentricity diffusion among planets takes place. 

Finally, in the planet-planet scattering scenario, the eccentricity grows at the dynamical timescale (comparable to the planetary orbital period), much faster than the secular timescale of the ZLK mechanism. Hence the presence of the moon does not affect the growth of eccentricity. On the other hand, close planetary encounters can eject the moons on wide orbits and perturb the innermost ones \citep{deienno14}. 

\subsection{Comparison to related works and impact on exomoon detectability}

Recent works have investigated survivability of exomoons of close-in giant planets. \citet{alvar17}, \citet{2019MNRAS.489.2313S} and \citet{tokad20} assume that the giant planet has successfully migrated close to star, and focus on the spin-orbit evolution of the coupled giant-star-moon system driven by mutual tides. They show that tides drive the migration of the lunar orbit over the timescale of ${>}1\,\rm Gyr$, possibly leading the lunar ejection.

Particularly, \citet{2019MNRAS.489.2313S} assume that once the moon reaches the planetary Hill radius, it will escape from the system, and model the post-escape dynamical evolution for $0.5 \myr$ using $N$-body simulations. They find that about $50\%$ of the moons survive the escape and become temporary planets on unstable orbits. This figure agrees with our results for low-mass moons (left column of Table~\ref{tab:sampout}), even though the mechanism leading to the instability is inherently different.  

We further show that by ${\sim} 10 \myr$, most of the moons have either collided or have been ejected from the system. Only ${\sim}1\%$ of the low-mass moons can remain as a stable planet, after their host planet has migrated close to the star. As a cautionary note, perturbations from the secondary star might increase the likelihood of the moon to collide with the primary star, so that such collisions could be increased in our scenario with respect to the single-star systems.

The detectability of exomoons around close-in giants was recently investigated by \citet{2020MNRAS.492.3499S}, who studied their secular migration due to time-dependent spin-orbit tidal coupling. They find that low mass ($m_\moon  / m_\planet < 10^{-4}$) moons are less likely to survive migration and they are also \revision{difficult to detect}. On the other hand, large exomoons migrate slower, and are more likely to be detected via transit-timing-variation and transit-duration-variation. 
Here we have shown that regardless of the lunar mass, moons are unlikely to survive the ZLK high-eccentricity migration of HJs. Thus, any future detection of an exomoon around a HJ would rule out its migration via this mechanism.

Our results also indicate that the lunar dynamical instability does not necessarily lead to the lunar ejection, but can lead the moon to tidally migrate towards stabler, tighter orbits around the planet (top-right panel of Figure~\ref{fig:outcome_example}), and even spiraling onto the planet. This interplay occurs when the perturbation from the secondary star onto the planet is barely strong enough to affect the lunar orbit, i.e. when the planetary Hill radius during one ZLK cycle is comparable to the lunar semimajor axis. The same mechanism could also occur in other scenarios, such as when the dynamical instability is triggered by the lunar outward migration, as considered in the works cited above. 
Spiral-in events could result in exorings around close-in giants, observable as additional dips in the planetary transit light curve \citep{2010Natur.468..943C,tusnski2011,kenworthy2015}. \revision{Debris around close-in giants could also outgas, fueling a plasma torus observable via high-resolution transmission spectroscopy \citep{oza19,gebek20}}

The collision of the moon and planets with the host star could also leave a debris disc of gas and dust around the star. This particular case was considered in detail by \citet{2019MNRAS.489.5119M}, who find that tidally detached exomoons on a highly eccentric orbit could evaporate, leaving an eccentric debris disc around the primary. Such a disc could explain the unusual dipping and secular dimming in the light curve of {\it KIC 8462852}, also known as Boyajian's Star \citep{boyaj16,wright16,metzger17,boyaj18,wyatt18}.

In the present work we have neglected the spin-orbit coupling term of tidal interactions. While spin-orbit coupling is an important factor when assessing the long term stability of exomoons, previous studies have shown that such evolution occurs over timescales much longer than than the timescales considered in our work \citep[${\lesssim}10\myr$ vs ${\gtrsim}1 \rm \, Gyr$, e.g.][]{alvar17,2019MNRAS.489.2313S}. Therefore, including the effect of spin would not alter our conclusions.

\section{Conclusions}\label{sec:con}

The first exomoon detection might occur in this decade. Besides the speculation that they might harbor life \citep{1997Natur.385..234W,heller13,heller14,martirod19}, the detection of exomoons can provide unique insights on planetary formation and evolution. 

In this work we have explored the role of exomoons in the high-eccentricity migration of HJs in the binary-ZLK scenario. Exomoons around Jupiters are not only unlikely to survive the migration of their host planet, but can even prevent the migration process by suppressing the ZLK oscillations induced by the secondary star. We term this effect as `moon shielding'.




The shielding effect is caused by apsidal precession induced by the moon on the planetary orbit around the primary star, which occurs on the ZLK timescale of the primary-planet-moon system. If this ZLK timescale is longer than that of the primary-secondary-Jupiter system, the planetary eccentricity can be freely excited. The periapsis of the planet shrinks until the moon becomes dynamically unstable, which can lead to a variety of outcomes, the most common of which is the collision of the moon with the primary star.

We sampled a slice of the initial parameter space that leads to the formation of a HJ in this scenario, and evolved such systems by including the presence of a moon around the planet. We evolved the systems using highly-accurate direct $N$-body integration, which included the effects of tides on each body, and relativistic precession.

In ${\sim}10\%$ of the systems, a massive ($0.1 \mjup$) moon is able to shield the planet efficiently from the perturbations of the secondary star. 
For less massive moons ($0.01 \mjup$), the percentage of efficiently shielded planets drops to $0.6\%$, in qualitative agreement with our semi-analytic predictions. In all other cases, the moon becomes dynamically unstable. 

In ${\sim}20\%$ of the times, the dynamical instability leads the moon to inspiral onto the planet. These kind of events can potentially form a system of rings around close-in giants, which would be detectable as additional dips in the planetary transit light curve \citep[e.g.][]{2018haex.bookE..35H}.

Between ${\sim}20$--$30\%$ of the exomoons collide immediately with the primary star, while the others will temporarily keep orbiting the primary as planets. In this latter case, the lunar orbit undergoes scatterings with the former host planet, until either one collides, gets ejected, or tidally circularizes around the primary. In total, ${\sim}30$--$35\%$ of the moons get ejected from the system and become  free-floating planets. 

A moon colliding with its parent star might leave an eccentric debris disc around the star. Such a disc could be at the origin of the anomalous lightcurve of {\it KIC 8462852} \citep{metzger17,2019MNRAS.489.5119M}, and even be observable in the near-infrared \citep[e.g.][]{jura03,2013Sci...342..218F}. The engulfment of massive exomoons could also explain the chemical dishomogeneity in binary systems \citep[e.g.][]{nagar20}.

Only about $1$--$6\%$ of the moon-turned-planets can become a stable planet around the primary, and about $2\%$ of the most massive moons undergo tidal decay and become hot Neptunes (Figure~\ref{fig:aemoon_post}). 

In the case that an exomoon will be detected around a close-in giant, this will be indicative of the migration mechanism of its host planet. Based on our population synthesis study and semi-analytic calculations, we can exclude that an exomoon could survive the migration of its host Jupiter in the binary-ZLK scenario.

\section*{Acknowledgements}
\revision{We thank the referee for his/her positive and constructive review of the manuscript}. The simulations were run on the CfCA Calculation Server at NAOJ. A.A.T. acknowledges support from JSPS KAKENHI Grant Numbers 17F17764 and 17H06360. A.A.T. also thanks the Center for Interdisciplinary Exploration and Research in Astrophysics at Northwestern University for its hospitality. A.S.H. thanks the Max Planck Society for support through a Max Planck Research Group.

\section*{Data availability}
The \texttt{TSUNAMI} code, the initial conditions and the simulation data underlying this article will be shared on reasonable request to the corresponding author.




\bibliographystyle{mnras}
\bibliography{literature,ms} 



\appendix
\section{Analytic estimates including short range forces}

\begin{figure}
\center
\includegraphics[width = 1\columnwidth]{{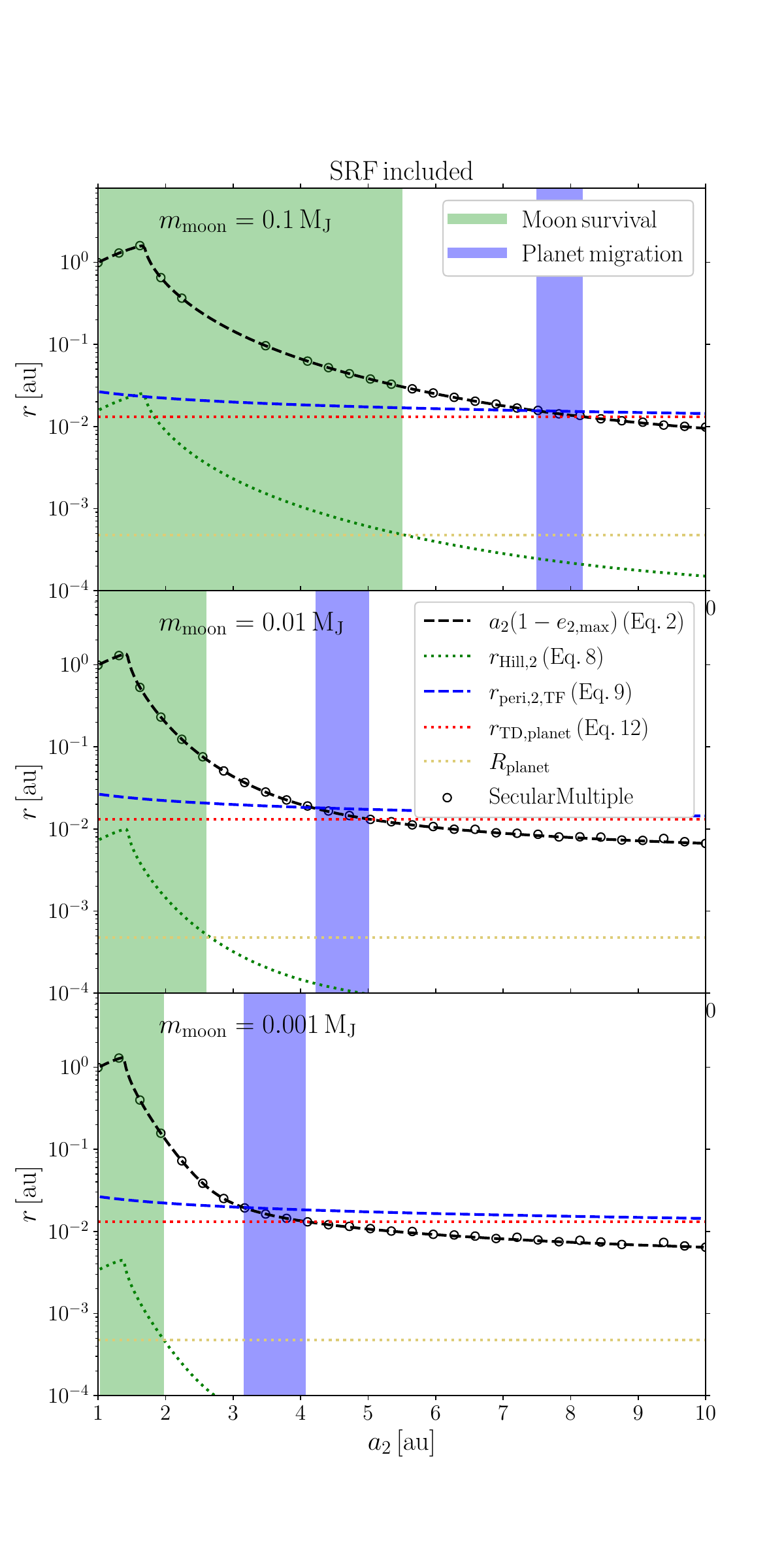}}
\caption{\small Similar to Figure~\ref{fig:semianalytic_noSRF}, here with the inclusion of additional short-range forces in the planet-moon orbit due to general relativistic corrections (1PN terms), and tidal bulges in the planet assuming an apsidal motion constant of $k_{\mathrm{AM,\,planet}}=0.19$, and a planetary radius of $R_{\mathrm{planet}}=1\,\rjup$.}
\label{fig:semianalytic_SRF}
\end{figure}

In Section~\ref{sec:an} we calculated the maximum eccentricity reached by the planet using Equation~\ref{eq:emaxmoon}, which considers the effect of the moon but neglects additional short-range forces such as those due to general relativistic corrections and tidal bulges. In Figure~\ref{fig:semianalytic_SRF}, we show a similar figure as Figure~\ref{fig:semianalytic_noSRF}, except that we included short-range forces in the planet-moon orbit due to the lowest-order post-Newtonian (PN) terms, as well as due to tidal bulges. These short-range forces give rise to additional apsidal motion, which tends to quench secular eccentricity excitation. Here, the semianalytic calculation is carried out by adding the relevant terms to the Hamiltonian in Equation~\ref{eq:emaxmoon} (e.g., \citealt{2007ApJ...669.1298F}), assuming a planetary apsidal motion constant of $k_{\mathrm{AM,\,planet}}=0.19$, and a planetary radius of $R_{\mathrm{planet}}=1\,\rjup$. We also include numerical results using \textsc{SecularMultiple} (black open circles), which confirm the validity of the semianalytic approach. 

With additional short-range forces included, the picture described in Section~\ref{sec:an} does not fundamentally change: moons around relatively close planets ($a_\planetprim \sim$ few $\au$) are effectively able to shield the planetary orbit, preventing migration of the planet. Much less massive moons, or moons around planets with larger $a_\planetprim$, have less shielding strength, but in this case, it is unlikely that they could survive the migration process.

\bsp	
\label{lastpage}
\end{document}